\documentclass[aps,prb,superscriptaddress,showkeys,showpacs,amsmath,amssymb,citeautoscript,reprint]{revtex4-2}
\usepackage[colorinlistoftodos]{todonotes}
\usepackage{soul}
\usepackage{multirow}
\usepackage{tabularx}
\usepackage{booktabs}
\usepackage{threeparttable}
\usepackage{xfrac}
\usepackage{bm}
\usepackage{graphicx}
\usepackage{rotating}
\usepackage{overpic}
\usepackage{subcaption}
\usepackage[colorlinks=true,linkcolor=blue,citecolor=red,urlcolor=blue]{hyperref}
\usepackage[capitalise]{cleveref}
\crefname{section}{Section}{Sections}
\usepackage{url}
\usepackage{orcidlink}
\usepackage{xspace}
\usepackage[shortlabels]{enumitem}
\setlist{noitemsep, leftmargin=*}
\def\C2N{Centre for Nanoscience and Nanotechnology (C2N) -- French National Centre for Scientific Research (CNRS), Paris, France}
\def\UPSaclay{Université Paris-Saclay, Paris, France}

\newcommand{\QE}{\textsc{Quantum ESPRESSO}\xspace}
\newcommand{\VASP}{Vienna Ab initio Simulation Package\xspace}
\newcommand{\BGW}{\textsc{BerkeleyGW}\xspace}
\newcommand{\Wannier}{\textsc{Wannier90}\xspace}
\newcommand{\GW}{$GW$\xspace}
\newcommand{\GoWo}{$G_{0}W_{0}$\xspace}
\newcommand{\GoWohse}{$G_{0}W_{0}$@HSE06\xspace}

\newcommand{\HSEscreen}{HSE$_\text{screen}$\xspace}
\newcommand{\HSEsolHFmix}{HSEsol$_\text{HFmix}$\xspace}
\newcommand{\InAsSb}{InAs$_x$Sb$_{1-x}$\xspace}

\newcommand{\suppinfo}{Supplemental Material (SM)~\cite{sm_insb_2025} 
    (see also references
    ~\cite{head_scikit-optimize_2024, rasmussen_gaussian_2008, jones_efficient_1998, srinivas_gaussian_2012, brennaman_insights_2023}
    therein)}

\begin{document}
\title{High-fidelity electronic structure and properties of InSb: G$_0$W$_0$ and Bayesian-optimized hybrid functionals and DFT+$U$ approaches}

\author{Ritwik Das\,\orcidlink{0000-0003-3073-0963}}
\email[Corresponding author:\ ]{ritwik.das@universite-paris-saclay.fr}
\homepage[\newline]{https://ritwikdas.gitlab.io}
\affiliation{\C2N}
\affiliation{\UPSaclay}

\author{Anne-Sophie Grimault-Jacquin\,\orcidlink{0009-0002-8966-0445}}
\affiliation{\C2N}
\affiliation{\UPSaclay}

\author{Frédéric Aniel\,\orcidlink{0009-0003-3620-8022}}
\affiliation{\C2N}
\affiliation{\UPSaclay}

\date{\today}

\begin{abstract}
  This study presents a refined approach to computing the electronic structure of indium antimonide (InSb) using advanced \textit{ab initio} techniques with the In and Sb $4d^{10}$ semicore electrons included in the valence states.
  These states are modeled using fully relativistic projector augmented waves (PAW) and optimized norm-conserving Vanderbilt (ONCV) pseudopotentials.
  However, standard Kohn-Sham density-functional theory (DFT) calculations with these pseudopotentials often produce non-physical band inversions and incorrect band gaps at the $\Gamma$-point due to $5p$-$4d$ repulsion and self-interaction errors (SIE).
  To resolve these issues, we apply a combination of hybrid Heyd-Scuseria-Ernzerhof (HSE) exchange-correlation (XC) functionals, many-body perturbation theory (MBPT) via quasiparticle $G_0W_0$, and DFT+U, significantly improving the accuracy of the band structure over previous studies.
  A Bayesian optimization framework is used to refine key parameters, including the inverse screening length ($\bm{\mu}$) and Hartree-Fock (HF) exchange fraction ($\bm{\alpha}$) in HSE-based XC functionals, as well as the Hubbard $\bm{U}$ parameters in DFT+U, leading to significantly improved band structure predictions.
  This approach yields highly precise band gaps, bulk moduli, effective masses, Luttinger parameters, valence bandwidth, and $4d$ band positions, achieving unprecedented agreement with experimental data.
  The resulting model resolves long-standing incomplete description of InSb's electronic band structure and provides a transferable computational framework for accurate electronic structure predictions across diverse material systems, offering valuable insights for future electronic, optoelectronic, energy, and quantum applications.
\end{abstract}

\keywords{Bayesian optimization, InSb, Electronic band structure, DFT, Hubbard, Hybrid functionals, DFT+U, GW, Effective mass, Luttinger parameter}

\maketitle

\section{Introduction}\label{sec:intro}

Indium antimonide (InSb), a key representative of zinc-blende (ZB) III-V semiconductors, crystallizes in a face-centered cubic (FCC) structure with space group \textit{F$\overline{4}$3m}.
It has attracted significant interest because of its exceptional electronic properties.
InSb has a direct band gap and it is the smallest band gap (0.17~$eV$ at 300~$K$, 0.23~$eV$ at 0~$K$) among all binary semiconductors, along with high electron mobility, small effective masses, long mean-free-path, large Landé g-factor, gate-tunable Rashba spin coefficients, and robust spin-orbit coupling (SOC)~\cite{mujica_high-pressure_2003, madelung_group_2002}.
These unique properties make InSb a foundational material for next-generation semiconductor and spintronic devices and for applications such as light-emitting diodes, thermal imaging, infrared detectors, magnetoresistive sensors, Majorana devices, and topological quantum computing~\cite{chen_strong_2021, dam_room-temperature_2012, jia_monolithic_2018, gul_ballistic_2024, frolov_topological_2020, op_het_veld_-plane_2020}.

However, despite the potential of InSb, accurately predicting its electronic band structure remains a formidable challenge primarily due to strong SOC and the influence of highly localized semicore $4d^{10}$ states.
Standard Kohn-Sham density-functional theory (DFT)~\cite{parr_density-functional_1995}, a powerful tool, suffers from well-known limitations when applied to InSb.
The lack of integer (derivative) discontinuity in the exchange-correlation (XC) energy ($E_{xc}$)~\cite{mori-sanchez_derivative_2014}, potential discontinuities while altering electron numbers and self-interaction errors (SIE)~\cite{cohen_insights_2008} lead to artificial stabilization of delocalized states, affecting the accuracy of electronic band structure resulting in incorrect band ordering and significant underestimation of the one-electron band gap compared to the experimental and quasiparticle (QP) ones~\cite{jones_density_1989, mori-sanchez_localization_2008}.

Recent advances in first-principles electronic-structure calculations have relied on diverse DFT codes, including plane-wave-based packages such as \VASP~\cite{hafner_ab-initio_2008}, \QE~\cite{giannozzi_advanced_2017}, and \textsc{Abinit}~\cite{gonze_abinit_2009}, as well as all-electron atomic-orbital codes like \textsc{WIEN2k}~\cite{blaha_wien2k_2020}. For III-V semiconductors such as InSb, accurate treatment of semicore states (In, Sb 4$d^{10}$) and inclusion of spin-orbit coupling (SOC) are crucial for predicting band ordering and $\Gamma$-point band gaps~\cite{malone_quasiparticle_2013}. Among pseudopotentials, the PAW method~\cite{blochl_projector_1994, torrent_implementation_2008} offers near all-electron accuracy; ultrasoft pseudopotentials (USPP)~\cite{vanderbilt_soft_1990, kresse_ultrasoft_1999} are efficient but less transferable; and optimized norm-conserving (ONCV) pseudopotentials~\cite{hamann_optimized_2013, schlipf_optimization_2015} provide high accuracy with slightly larger cutoff requirements. Benchmarking studies~\cite{lejaeghere_reproducibility_2016, van_setten_pseudodojo_2018} show that PAW and ONCV outperform USPP in structural and spectral predictions. For InSb--with its narrow gap, strong SOC, and 4$d$ repulsion--fully relativistic ONCV or PAW pseudopotentials ensure better accuracy. We use \QE for its full support of ONCV, relativistic corrections, and compatibility with our Bayesian optimization workflow.

A persistent issue in prior studies is the inconsistent treatment of semicore $4d^{10}$ states. Kim \textit{et al.}~\cite{kim_towards_2010} investigated InSb using both \VASP and \textsc{WIEN2k}. In their \VASP-based calculations, Sb $4d$ electrons were treated as core, yielding band gaps of 0.28, 0.24, and 0.35~eV for HSE06, HSE$_\mathrm{bgfit}$ (with manually tuned $\mu$), and $G_0W_0^{\text{TC-TC}}$ (test-charge--test-charge), respectively. By contrast, their all-electron \textsc{WIEN2k} calculation with MBJLDA treated $4d$ electrons as valence and yielded a larger gap of 0.26~eV. Although  HSE$_\mathrm{bgfit}$ method closely approximate the experimental band gap of 0.235~eV, this plane-wave pseudopotential-based result lack a complete and accurate description of semicore valence states, and result in discrepancies in band dispersion and effective masses. These differences highlight the needs of explicitly treating semicore states when modeling narrow-gap semiconductors.

To overcome the limitations of obtaining accurate band structure with $4d$ valence states, we employ a combination of state-of-the-art \textit{ab initio} methods---including Hartree-Fock (HF) and DFT-based hybrid Heyd-Scuseria-Ernzerhof (HSE) functionals, many-body perturbation theory (MBPT) based QP-\GW, and density-functional perturbation theory (DFPT) based DFT+U. This methodological synergy addresses the $5p$-$4d$ level repulsion and self-interaction errors (SIE) exacerbated by strong SOC, which have historically hindered accurate modeling of InSb. In this work, we explicitly treat the In and Sb $4d^{10}$ states as valence electrons using fully relativistic PAW and ONCV pseudopotentials. Our approach delivers the most accurate \GoWo band structure reported to date, surpassing even the recent efforts by Gant \textit{et al.} (2022)~\cite{gant_optimally_2022}, which failed to reproduce correct band structure and band gaps of InSb.

Moreover, in this work, we introduce a Bayesian optimization framework integrated with \QE~\cite{giannozzi_advanced_2017} to systematically refine key functional parameters---including the inverse screening length ($\bm{\mu}$) and exchange fraction ($\bm{\alpha}$) in HSE XC-functionals, as well as the Hubbard $\bm{U}$ parameters in DFT+U. Traditional empirical fitting---based on reproducing experimental band gaps~\cite{kim_towards_2010}---lacks first-principles rigor and fails to generalize across chemically distinct systems or novel materials where no experimental data are available~\cite{zhao_structure_2022, gebauer_oxygen_2023}. Alternatively, for Hubbard $U$ parameters, linear-response methods~\cite{cococcioni_linear_2005}, while formally grounded in DFT, are computationally intensive and sensitive to the choice of localized projectors, and in the case of InSb, yield inaccurate band gaps and valence band curvatures (see Sec.~\ref{sec:results}). In contrast, our approach leverages Gaussian process regression to efficiently explore the parameter space and iteratively minimize discrepancies with a high-level \GoWo reference, requiring a minimal number of DFT evaluations. This data-driven strategy supports multi-parameter optimization, accelerates convergence, and is tightly coupled with \QE, selected for its robust support of full-relativistic PAW and ONCV pseudopotentials, open-source availability, and seamless integration with external workflows, including our Python-based optimization engine.

Unlike earlier studies focused primarily on the band gap~\cite{kim_towards_2010, gant_optimally_2022}, our methodology provides a comprehensive and transferable framework for computing the full band structure of InSb---including spin-orbit splittings, effective masses, and Luttinger parameters---while resolving discrepancies related to semicore state placement and band ordering. It offers a tractable alternative to many-body methods, delivering high-fidelity results at a fraction of the computational cost, and is readily extensible to III-V and II-VI semiconductors relevant for infrared optoelectronics, spintronic transport, and quantum information applications.

\section{Computational Details}\label{sec:comp_details}

\begin{figure}[!htbp]
  \centering
  \includegraphics[width=\columnwidth]{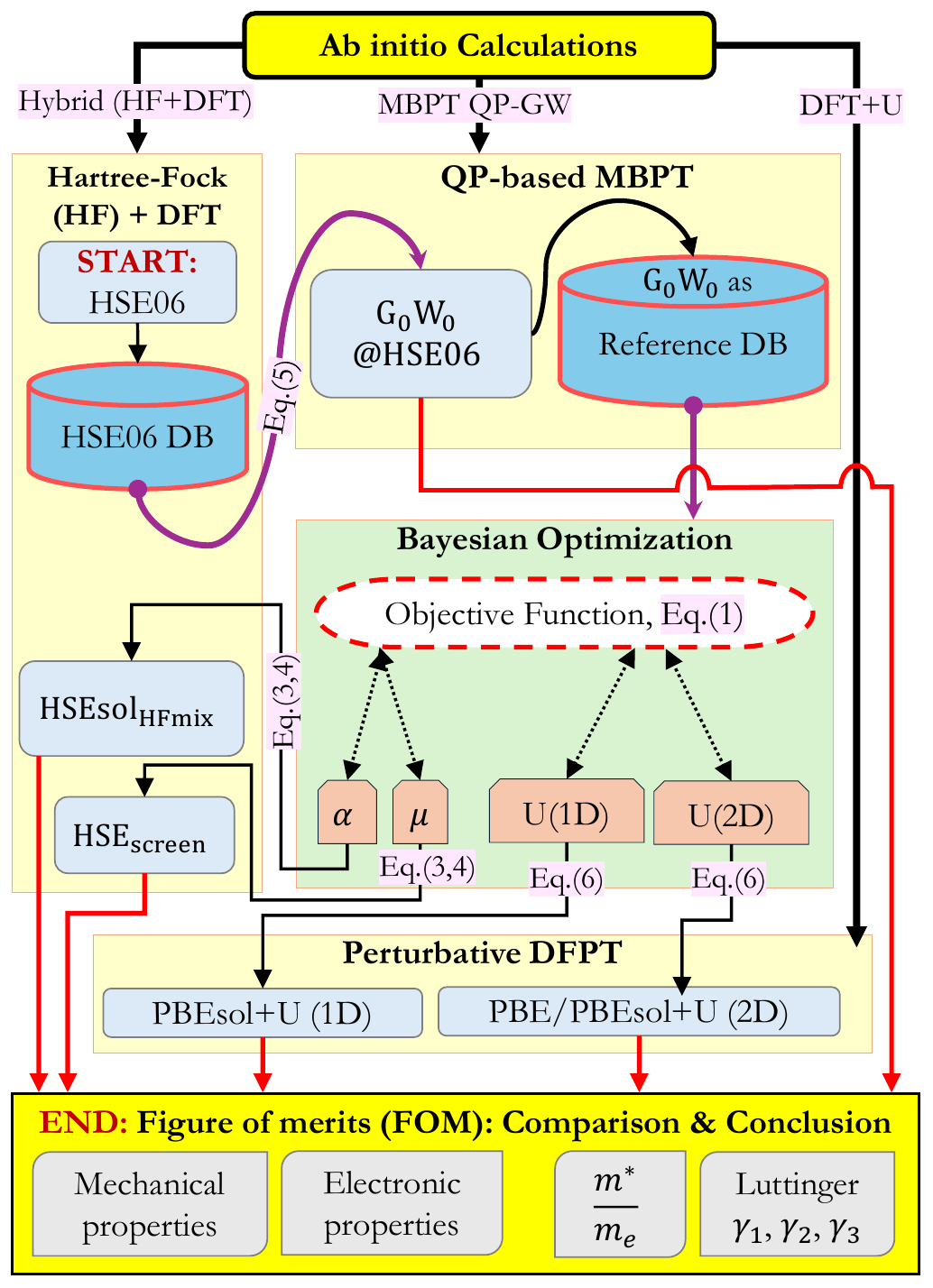}
  \caption{
    \textbf{Schematic workflow of the computational methodology.}
    \textbf{Legend:}
    \textbf{``DB''} (blue cylinder with red outline) refers to an internal database storing HSE and \GoWo reference data.
    \textbf{Solid arrows} denote standard computational flow;
    \textbf{purple arrows} represent parameter or data retrieval from the DB;
    \textbf{Red arrows} indicate data flow from prior \textit{ab initio} calculations, which are then fed forward for use in generating final figures of merit (FOM) in the evaluation block.
    \textbf{Rounded-cornered rectangles} represent individual \textit{ab initio} calculations;
    \textbf{Light brown rectangles} represent optimized parameters ($\bm{\alpha}$, $\bm{\mu}$, $\bm{U}$(1D,2D)) extracted from Bayesian optimization (BO).
    \newline
    \textbf{Color code:}
    \textit{pale yellow}---first-principles calculations;
    \textit{light green}---BO block;
    \textit{yellow}---evaluation and comparison stage.
    \newline
    The diagram summarizes five main components:
    \newline
    (i) \textbf{\textit{Hybrid-DFT block}} begins with \texttt{HSE06} and also contains its two variants (\texttt{\HSEscreen}, \texttt{\HSEsolHFmix});
    (ii) \textbf{\textit{Quasiparticle (QP) \GoWo calculations}}, based on HSE06 wavefunctions, serve as the reference to guide BO of hybrid and DFT+$U$ parameters;
    (iii) \textbf{\textit{Bayesian Optimization}} refines the parameters $\bm{\mu}$, $\bm{\alpha}$, and $\bm{U_{\mathrm{eff}}}$;
    (iv) \textbf{\textit{DFT+$U$ calculations}} use the optimized $\bm{U}$ parameters;
    (v) \textbf{\textit{Final predictions}} yield key figures of merit: structural, electronic, and transport properties.
  }
  \label{fig:workflow}
\end{figure}

\subsection{Pseudopotential Generation}\label{subsec:pseudo_gen}

In our \textit{ab initio} calculations, we employed both Projector Augmented Wave (PAW)~\cite{blochl_projector_1994} and Optimized Norm-Conserving Vanderbilt (ONCV)~\cite{hamann_optimized_2013, schlipf_optimization_2015} pseudopotentials to balance accuracy and computational efficiency.
PAW pseudopotentials were generated using the \textit{ld1.x} utility of the \QE distribution~\cite{giannozzi_advanced_2017}.
These customized potentials balance the computational efficiency of pseudopotentials and the accuracy of all-electron calculations, particularly for elements with strong core-valence interactions.
For hybrid-HSE, \GoWo, and DFT+$U$ calculations, we utilized ONCV pseudopotentials generated using the ONCVPSP code with the Kleinman-Bylander (KB) approach for optimization~\cite{schlipf_optimization_2015}.
ONCV pseudopotentials are exceptionally advantageous due to their ability to preserve norm conservation while offering improved transferability and smoother pseudo-wave functions compared to traditional norm-conserving pseudopotentials.

Given the low position of In and Sb in Mendeleev's periodic table, it is essential to account for spin-orbit coupling (SOC) in DFT calculations.
We implemented fully relativistic (FR) SOC in all pseudopotentials.
The Generalized Gradient Approximation (GGA)-based Perdew-Burke-Ernzerhof (PBE) and its solid-state variant PBEsol XC functionals were incorporated using the LibXC library~\cite{lehtola_recent_2018}.

\begin{table*}[!tbp]
  \caption{Comparison of pseudopotential (PP) parameters for Indium (In) and Antimony (Sb) using PAW and ONCV methods with PBE and PBEsol exchange-correlation (XC) functionals. For each PP type, the table lists the valence electron configuration (`\texttt{Valence}'), the maximum core radius (\texttt{`$R_{core}$' in a.u.}), the form of the local potential~$^a$, the nonlocal projector components (with the number of projectors per angular momentum channel in parentheses), and the minimum wavefunction and charge density energy cutoffs (\texttt{`$E_{\text{cut}}$' in Rydberg, Ry}).}
  \label{tab:pp_conf}
  \begin{ruledtabular}
    \begin{tabularx}{\textwidth}{@{} l l c c c c c c c @{}}
      \multirow{2}{*}{\parbox{10mm}{\raggedright \textbf{PP Type}}}                           &
      \multirow{2}{*}{\parbox{11mm}{\raggedright \textbf{XC Func.}}}                          &
      \multirow{2}{*}{\parbox{15mm}{\raggedright \textbf{Element}}}                           &
      \multirow{2}{*}{\parbox{19mm}{\centering \textbf{Valence Config.}}}                     &
      \multirow{2}{*}{\parbox{20mm}{\centering \textbf{Maximum R$_{core}$ (a.u.)}}}           &
      \multirow{2}{*}{\parbox{18mm}{\centering \textbf{Local Potential}~\footnotemark[1]}}    &
      \multirow{2}{*}{\parbox{32mm}{\centering \textbf{Nonlocal Proje- ctors (No. per $l$)}}} &
      \multirow{2}{*}{\parbox{16mm}{\centering \textbf{Wavefunc. E$_{cut}$ (Ry)}}}            &
      \multirow{2}{*}{\parbox{30mm}{\centering \textbf{Charge Density E$_{cut}$ (Ry)}}}                                                                                                                    \\
                                                                                              &                         &    &                         &      &
                                                                                              &                                                                                                            \\
      \midrule
      \multirow{2}{*}{\centering PAW}                                                         & \multirow{2}{*}{PBE}    & In & 4d$^{10}$ 5s$^2$ 5p$^1$ & 1.90 & AE potential & s(1), p(2), d(2) & 41 & 163 \\
                                                                                              &                         & Sb & 4d$^{10}$ 5s$^2$ 5p$^3$ & 2.10 & AE potential & s(1), p(2), d(2) & 43 & 171 \\
      \multirow{2}{*}{\centering ONCV}                                                        & \multirow{2}{*}{PBE}    & In & 4d$^{10}$ 5s$^2$ 5p$^1$ & 2.35 & AE potential & s(2), p(2), d(2) & 65 & 262 \\
                                                                                              &                         & Sb & 4d$^{10}$ 5s$^2$ 5p$^3$ & 2.40 & AE potential & s(2), p(2), d(2) & 73 & 290 \\
      \multirow{2}{*}{\centering ONCV}                                                        & \multirow{2}{*}{PBEsol} & In & 4d$^{10}$ 5s$^2$ 5p$^1$ & 2.35 & AE potential & s(2), p(2), d(2) & 66 & 265 \\
                                                                                              &                         & Sb & 4d$^{10}$ 5s$^2$ 5p$^3$ & 2.40 & AE potential & s(2), p(2), d(2) & 75 & 295 \\
    \end{tabularx}
  \end{ruledtabular}
  \footnotetext[1]{In ONCV pseudopotentials, the local potential is constructed as a smooth polynomial continuation of the all-electron potential, rather than being derived from a specific angular momentum channel.}
\end{table*}

\cref{tab:pp_conf} provides a comprehensive overview of the pseudopotential parameters for In ($4d^{10} 5s^{2} 5p^{1}$) and Sb ($4d^{10} 5s^{2} 5p^{3}$), treating their $4d^{10}$ semicore electrons as valence states.
It includes each element's pseudopotential type, XC functional, valence electron configuration (`\texttt{Valence}'), maximum core radius (\texttt{`$R_{core}$' in a.u.}), local potential projector, nonlocal components, and the plane-wave and charge-density cutoff energies (\texttt{`$E_{cut}$'  in Rydberg, $Ry$}).
The table facilitates a comparative understanding of the pseudopotential parameters for In and Sb, highlighting differences in local and nonlocal potential treatments across PAW and ONCV methods with PBE and PBEsol XC functionals.

\subsection{\textit{Ab initio} DFT, DFPT and MBPT-based \GoWo}\label{subsec:abinitio_calculs}

\textit{Ab initio} structure relaxations and electronic band structure computations were performed within the DFT framework using a Plane-Wave (PW) basis set and pseudopotentials-based method as implemented in \QE.
In this work, all electronic structure and energy gap calculations were performed at the 300~K experimental lattice constant of 6.479~\AA, except for the determination of structural parameters (e.g., relaxed lattice constants and bulk moduli), which were obtained by fully relaxing the lattice geometry.
The quasiparticle \GoWo calculations are done using the \BGW distribution~\cite{barker_spinor_2022}.

Brillouin-zone (BZ) integrations were executed using $\Gamma$-centred $\bm{k}$-point meshes.
Self-consistent field (SCF) calculations for determining the ground state charge density, equilibrium lattice constants and bulk moduli, an $8\!\times\!8\!\times\!8$ $\bm{k}$-point grid was used, yielding 29 irreducible $\bm{k}$-points in the first BZ. In SCF calculations, Gaussian smearing with a smearing width of 0.005 Rydberg is used.
For non-self-consistent field (NSCF) calculations, a denser $16\!\times\!16\!\times\!16$ $\bm{k}$-point grid was employed to obtain more accurate eigenvalues and improve the resolution of the band, particularly near the Fermi level.
No symmetry operations are used in all calculations to ensure accurate band structure results.
We also extended this approach by not considering the time-reversal symmetry $\left(\bm{k}\rightarrow-\bm{k}\right)$, thereby treating the full $8\!\times\!8\!\times\!8$ BZ mesh to encompass all 512 $\bm{k}$-points as distinct and desymmetrizing the charge density, enhancing our computational results' precision.
The band structures $\mathcal{E}(\bm{k})$ were computed on a discrete $\bm{k}$ mesh along high-symmetry directions, i.e., from the BZ center $\Gamma$ (0,~0,~0) to $X$ (0.5,~0.0,~0.5), $W$ (0.50,~0.25,~0.75), $L$ (0.5,~0.5,~0.5), $\Gamma$, and $K$ (0.375,~0.375,~0.750) in crystal units.
A fixed occupation is used to manage electron occupation in NSCF and band structure calculations.

For hybrid-HSE functional based calculations, the non-local Fock exchange term was evaluated on a $\Gamma$-centered $4\!\times\!4\!\times\!4$ $\bm{q}$-point mesh, which ensures adequate convergence of the screened exchange interaction in reciprocal space.
Eigenvalues of the hybrid HSE and \GoWo band structures are interpolated using the \Wannier code~\cite{pizzi_wannier90_2020}.
Given the similar orbital character of the states near the band gap of InSb, we focus on Wannierizing only the eight highest occupied and six lowest unoccupied bands around the band edges, employing $sp^3$ initial projections.
SOC corrections are applied to the interpolated bands for each eigen energy $\mathcal{E}_{n\bm{k}}$ and are subsequently interpolated using maximally localized Wannier functions (MLWFs), following the method described by Malone and Cohen~\cite{malone_quasiparticle_2013}.

The linear-response Hubbard $U$ values were computed using the \texttt{hp.x}~\cite{timrov_hp_2022} module of \QE, which implements linear response theory (LRT) within DFPT. It computes the response of localized atomic occupations to on-site potential shifts, yielding a formally \textit{ab initio} estimate of $U$. In this approach, the interacting ($\chi$) and noninteracting ($\chi^0$) susceptibilities are evaluated self-consistently in a supercell using DFPT, without relying on total-energy derivatives.

\subsection{Bayesian-Optimized Methodological Workflow}\label{subsec:workflow}

Recent advances in machine learning for electronic-structure theory have begun to reshape how DFT and its core components are developed. Notably, \textsc{Microsoft Research}'s ``AI for Science'' recently introduced the \textsc{Skala} framework, an accurate and scalable deep-learning surrogate for exchange-correlation (XC) functionals trained on high-level quantum chemical data~\cite{luise_accurate_2025}. While \textsc{Skala} represents a significant step toward end-to-end learned functionals, our work focuses on a complementary direction: the optimization of existing DFT-based approaches through Bayesian surrogate modeling~\cite{das_bmach_2024}. Specifically, we present an advanced computational workflow that integrates Hartree-Fock (HF) and DFT-based hybrid-HSE XC functionals, Hubbard-corrected DFT+U methods, many-body perturbation theory (MBPT)-based \GoWo calculations, and machine-learning-based Bayesian optimization techniques~\cite{shahriari_taking_2016, yu_machine_2020, das_bmach_2024}. The essential computational steps are summarized below:

\begin{enumerate}[1.]
  \item Hybrid (HF+DFT) calculations using HSE06.
  \item Many Body Perturbation Theory (MPBT) based single-shot quasiparticle (QP) \GoWo calculations from HSE06 ``starting-point'' (reported as \GoWohse).
  \item Bayesian optimization~\cite{shahriari_taking_2016, yu_machine_2020, das_bmach_2024}, utilizing \GoWohse as the reference, to optimize:
        \begin{enumerate}[(a)]
          \item screening parameter ($\bm{\mu}$) in \HSEscreen,
          \item exchange fraction ($\bm{\alpha}$) in \HSEsolHFmix,
          \item effective Hubbard $\bm{U}$ parameters for DFT+U.
        \end{enumerate}
  \item Modified hybrid-HSE calculations (\HSEscreen and \HSEsolHFmix) using parameters ($\bm{\mu}$ and $\bm{\alpha}$) optimized in the previous BO step.
  \item DFT+$U$ calculations with $\bm{U}$ optimized in step 3c.
\end{enumerate}

\Cref{fig:workflow} visually consolidates these steps into an integrated pipeline, highlighting how Bayesian optimization connects \GoWo, hybrid-DFT, and DFT+$U$ calculations to refine key parameters and improve predictive accuracy. All functionals and methods employed are detailed in \cref{subsec:functionals}.

The self-developed Bayesian optimization algorithm employing a Gaussian Process model with a Matern-5/2 kernel combined with a WhiteKernel to account for observational noise was used to optimize parameters and establish agreement with \GoWohse results. The optimization process was fully automated, dynamically adjusting acquisition functions to balance exploration and exploitation. For Kernel and model details see Sec.~S2 of the of the \suppinfo.
The optimization is captured by the objective function $f(x)$, defined in \cref{eq:obj_func}.

\begin{equation}\label{eq:obj_func}
  f\left(x\right) = a_1 \left(E_0^{Ref.}-E_0^{Targ.}\right)^2 + a_2 \left(\Delta\mathcal{E}_{BZ}\right)^2
\end{equation}

Here, $E_0$ denotes the band gap, and $\Delta\mathcal{E}_{BZ}$, the eigenvalues' variation across the 3D Brillouin Zone (BZ) as defined in \cite{huhn_one-hundred-three_2017}, is shown in \cref{eq:del_band}.

\begin{equation}\label{eq:del_band}
  \Delta\mathcal{E}_{BZ}=\sqrt{\frac{1}{N_E}\sum_{i=1}^{N_k}\ \sum_{j=1}^{N_b}\left(\mathcal{E}_{Ref.}^j\left[k_i\right]-\mathcal{E}_{Targ.}^j\left[k_i\right]\right)^2}
\end{equation}

The variable $x$ in $f(x)$ represents the parameters being optimized ($\bm{\mu}$, $\bm{\alpha}$, or $\bm{U}$), aligning the reference (\GoWohse as $Ref.$) and target ($HSE_{screen}$, $HSEsol_{HFmix}$, $DFT$+$U$ as $Targ.$) calculations.

The optimization variables were treated as continuous parameters, each defined over a physically meaningful search space (typically $[-10, 10]$~eV), distinct from the energy window used to evaluate band eigenvalues. This continuous-domain treatment enables fine-grained optimization using Gaussian Process surrogates.

The weights $a_1$ and $a_2$ in the objective function act as hyperparameters to balance the influence of the band gap and band dispersion curvature in the optimization process. We adopted $a_1 = 0.75$ and $a_2 = 0.25$, prioritizing the band gap due to its critical role in determining optical and transport properties, while still ensuring reasonable agreement in the shape of the overall band structure. This adjustment is rationalized by the sensitivity of specific atomic orbital contributions: increasing the weight of the band gap term ($a_1$) tends to make the optimized $U$ values for Sb-$5p$ states less negative, while those for In-$5p$ states remain largely unaffected.
Since the top of the valence band is primarily dominated by Sb $p$-orbitals, correcting these states plays a particularly important role in accurately opening the gap. Our weighting strategy was manually tuned to reflect these physical considerations.
Notably, this weighting scheme aligns with that used by Yu \textit{et al.}~\cite{yu_machine_2020}, who adopted the same values to emphasize gap accuracy in their $U$ optimization. While our implementation differs in material and technical aspects--such as the reference method, orbital targets, acquisition function and BO kernel--the underlying rationale remais consistent, supporting the robustness and generality of this approach across material systems.

In the Bayesian optimization, band eigenvalues were evaluated within the energy range of $[-10, 10]$~eV, with the valence band maximum (VBM) set to 0~eV. Within this window, no band entanglement was observed. In cases where entanglement did occur in reference band structures (e.g., \GoWo and hybrid functional calculations), it was resolved using Wannier interpolation, which projects entangled states onto localized orbitals to produce smooth, disentangled bands. Bands below $-10$~eV, corresponding mainly to deep core states, were excluded from the optimization as our primary focus was on the band gap and valence/conduction dispersions near the Fermi level.

For DFT+$U$ corrections, two configurations were considered, where the terms \textbf{1D} and \textbf{2D} refer strictly to the dimensionality of the Bayesian optimization variable space--\emph{not} to one- or two-dimensional physical Hubbard models. These configurations are defined by the number of orbital-specific $U$ parameters optimized:
(i)~`\textbf{1D}' --- a single-parameter optimization in which $U$ is applied only to the Sb-$5p$ orbitals, and
(ii)~`\textbf{2D}' --- a two-parameter optimization where $U$ values for both In-$5p$ and Sb-$5p$ orbitals are simultaneously optimized.

\subsection{Functionals \& Methods: GW, HSE, DFT+U}\label{subsec:functionals}

\subsubsection{Hybrid functional formulation and HSE}\label{subsubsec:hybrid_formulation}

Hybrid functionals partition the Coulomb interaction into short-range (SR) and long-range (LR) components to efficiently capture both short- and long-range exchange interactions.
The splitting of the Coulomb potential in hybrid functionals like \textit{HSE} is achieved using the error function in \cref{eq:coulomb_split_sr_lr}~\cite{henderson_can_2009}.

\begin{align}\label{eq:coulomb_split_sr_lr}
  \frac{1}{r} = \frac{\bm{\alpha} + \bm{\beta} \cdot \text{erf}(\bm{\mu} r)}{r} + \frac{1 - [\bm{\alpha} + \bm{\beta} \cdot \text{erf}(\bm{\mu} r)]}{r}
\end{align}

In this equation, $\bm{\alpha}$ controls the amount of Hartree-Fock (HF) exchange in the short range, while $\bm{\alpha\!+\!\beta}$ controls the exact exchange in the long range.
The screening parameter ($\bm{\mu}$) governs the range-separation, smoothly transitioning from short-range to long-range interactions~\cite{henderson_can_2009}.

For hybrid functionals like \textit{HSE}, $\bm{\beta}$=$0$, simplifying the long-range part.

The total exchange-correlation energy in the \textit{HSE} functional is then defined as in \cref{eq:hse_energy}.

\begin{align}\label{eq:hse_energy}
  \mathcal{E}_{xc}^{HSE} = {} & \bm{\alpha} \mathcal{E}_x^{HF,SR}(\bm{\mu}) + (1-\bm{\alpha})\mathcal{E}_x^{PBE,SR}(\bm{\mu}) \nonumber \\
                              & + \mathcal{E}_x^{PBE,LR}(\bm{\mu}) + \mathcal{E}_c^{PBE}
\end{align}
where \textit{HF,SR} and \textit{PBE,LR} denote HF-type SR and PBE-type LR exchange, respectively.

\vspace{3mm}{\centering \textbf{\textit{1.a. HSE06}}\par}\label{subsubsec:hse06}

The \textit{HSE06} implementation is a specific variant of the general \textit{HSE} functional.
In \textit{HSE06}, the parameters are set to $\bm{\alpha}$=0.25 and $\bm{\mu}$ = 0.106~$bohr^{-1}$~\cite{heyd_erratum_2006}.
These values are used in the general \textit{HSE} equation (Eq.~\ref{eq:hse_energy}), meaning that 25\% of the short-range HF exchange is included, while the short-range interactions beyond $2/\bm{\mu} \approx 18.86$~\AA\ becomes negligible.

\Cref{tab:hse_variants} summarizes the key \textit{HSE} functional variants used in this work, with their corresponding $\bm{\alpha}$, $\bm{\beta}$, and $\bm{\mu}$ values as parameterized in \cref{eq:coulomb_split_sr_lr,eq:hse_energy}.

This table presents the functional variants with their associated parameters $\bm{\alpha}$, $\bm{\beta}$, and $\bm{\mu}$.
In HSE and its derivatives, $\bm{\beta}$=0, but future functionals may adjust this to control the amount of long-range exact exchange.

\begin{table}[!tbp]
  \caption{Key \textit{HSE} functional variants used in this work.
    The parameters $\bm{\alpha}$ (exchange fraction), $\bm{\beta}$ (long-range exchange mixing), and $\bm{\mu}$ (range-separation screening parameter) are defined in \cref{eq:coulomb_split_sr_lr,eq:hse_energy}.
    In \textit{HSE06} and all other \textit{HSE} derivatives, $\bm{\beta}$=0, as no long-range Hartree-Fock exchange is included.}
  \label{tab:hse_variants}
  \begin{ruledtabular}
    \begin{tabularx}{\columnwidth}{@{} l c c c @{}}
      \textbf{Hybrid Functional}                       &
      \textbf{$\bm{\alpha}$}                           &
      \textbf{$\bm{\beta}$}                            &
      \textbf{$\bm{\mu}$ (bohr$^{-1}$)}                                   \\
      \midrule
      HSE06~(\cref{subsubsec:hse06}.a)                 & 0.25 & 0 & 0.106 \\
      \HSEscreen~(\cref{subsubsec:optimized_hse}.b1)   & 0.25 & 0 & 0.095 \\
      HSEsol~(HSE06's solid variant)                   & 0.25 & 0 & 0.106 \\
      \HSEsolHFmix~(\cref{subsubsec:optimized_hse}.b2) & 0.30 & 0 & 0.106 \\
    \end{tabularx}
  \end{ruledtabular}
\end{table}

\vspace{3mm}{\centering \textbf{\textit{1.b. $G_0W_0$-data-driven HSE}}\par}\label{subsubsec:optimized_hse}
In range-separated hybrid functionals, the screening parameter ($\bm{\mu}$) is often adjusted to improve the comparison with experimental properties, such as band gaps~\cite{henderson_can_2009, kim_towards_2010}.
We have further advanced this approach by optimizing both the screening parameter $\bm{\mu}$ and the exchange fraction ($\bm{\alpha}$) using the Bayesian optimization (see \cref{subsec:workflow}).
Two such optimizations explored in this work are \HSEscreen and \HSEsolHFmix.
The optimized values of $\bm{\mu}$ and $\bm{\alpha}$ are indicated in parentheses in the method names reported in \cref{tab:hse_variants}.

In both cases, \GoWohse serves as the reference for optimization.

\vspace{1mm}\textbf{\textit{1.b.1. \HSEscreen:}}
At a fixed exchange fraction of $\bm{\alpha} = 0.25$, Bayesian optimization yields an optimal screening parameter of $\bm{\mu} = 0.095~\text{bohr}^{-1}$. This adjustment improves the electronic band gap prediction by 56~meV compared to standard HSE06 (see \cref{tab:mechanical_electronic}), as the default screening length in HSE06 is suboptimal for this system.

\vspace{1mm}\textbf{\textit{1.b.2. \HSEsolHFmix:}}
Here the screening parameter is fixed at $\bm{\mu}$ = 0.106~$bohr^{-1}$, while the exchange fraction is optimized to $\bm{\alpha}$=0.30.
This increases the proportion of HF exchange in the short-range region, making this optimized XC functional better suited for larger and accurate band gap prediction.

\subsubsection{Quasiparticle \GoWo Calculations (Used as Reference)}\label{subsubsec:g0w0}

The many-body perturbation theory (MBPT)-based single-shot quasiparticle (\GoWo) calculations are performed within the Green's function-based \GW approximation~\cite{zhu_quasiparticle_1991}. In principle, the \GW self-energy $\Sigma = iGW$ should be determined self-consistently; however, due to its high computational cost, we adopt the widely used single-shot \GoWo scheme. (See Sec.~S1 of SM, for the formal expressions of Dyson's equation, Green's function, and QP energy corrections.)

In this perturbative scheme, the QP energies ($\mathcal{E}_{n\bm{k}}^{\mathrm{QP}}$) are computed as first-order corrections to mean-field DFT eigenvalues ($\mathcal{E}_{n\bm{k}}^{\text{DFT}}$):
\begin{equation}\label{eq:qp_energy_main}
  \mathcal{E}_{n\bm{k}}^{\mathrm{QP}} = \mathcal{E}_{n\bm{k}}^{\text{DFT}} + \langle \psi_{n\bm{k}} | \Sigma(\mathcal{E}_{n\bm{k}}^{\text{DFT}}) - V_{xc} | \psi_{n\bm{k}} \rangle
\end{equation}

We note that the \GW method, being a perturbative quasiparticle (QP) correction, yields accurate excitation energies but does not provide total energies.

\GoWo calculations are performed starting from HSE06 eigenstates (denoted as \GoWohse in this work), using norm-conserving ONCV pseudopotentials. Spin-orbit coupling (SOC) effects are included self-consistently both in the construction of the dielectric screening $\bm{\varepsilon^{-1}}$ and in the evaluation of the self-energy $\bm{\Sigma}$. The frequency dependence of $\bm{\varepsilon^{-1}}$ is modeled using the Godb-Needs plasmon pole (GPP)~\cite{godby_metal-insulator_1989} approximation. To ensure numerical convergence, we include 500 and 1200 empty bands for the computation of the polarizability $\bm{\chi}$ and the Coulomb-hole self-energy contribution $\bm{\Sigma}_{\mathrm{CH}}$, respectively.

\subsubsection{Hubbard DFT+$U$ calculations}\label{subsubsec:dftu}

After employing the computationally demanding \GoWo and HSE methods, we transitioned to DFPT-based Hubbard-corrected DFT+$U$ calculations, which offer a more efficient alternative while effectively addressing self-interaction errors (SIE).
Using PBE and PBEsol XC functionals, the Hubbard correction introduces piecewise linearity in the energy functional, removing SIE within the Hubbard manifold and maintaining accuracy in vital electronic properties.

In our DFT+$U$-$J$ approach, we selectively apply the Hubbard correction to the $5p$-orbital manifolds while treating other delocalized states at the standard DFT level. Within the Dudarev formulation of DFT+U, the effective Hubbard $\bm{U}$ is defined as $\bm{U_{eff}}\! = \!\bm{U}\!-\!\bm{J}$, where $\bm{U}$ and $\bm{J}$ represent the on-site Coulomb repulsion and exchange interaction, respectively~\cite{shishkin_dftu_2019}. The total DFT+$U$-$J$ energy ($\mathcal{E}_{tot}$) is given by \cref{eq:hub_energy}.

\begin{equation}\label{eq:hub_energy}
  \mathcal{E}_{tot}=\mathcal{E}_{DFT}+\frac{U-J}{2}\sum_{\sigma}{\ n_{m,\sigma}-n_{m,\sigma}^2}
\end{equation}

where $n$ is the atomic-orbital occupation number, $m$ is the orbital momentum, and $\sigma$ is a spin index.

$\bm{U_{eff}}$ is represented as an n-dimensional vector $\vec{\bm{U}}$=$\left[U^1,U^2,\ldots,U^n\right]$ applied to different atomic species and orbitals ($U^n$). To determine optimal $\bm{U_{eff}}$ parameters, Bayesian optimization model as described in \cref{subsec:workflow} is used by minimizing the objective function in \cref{eq:obj_func}, with $x$=$\vec{\bm{U}}$, and $Tar.$ (target)=$DFT$+$U$.

For InSb's delocalized In-$5p$ and Sb-$5p$ states, negative $\bm{U_{eff}}$ values are used due to GGA's overestimation of the exchange-correlation hole~\cite{shishkin_dftu_2019}. Negative $\bm{U_{eff}}$ is theoretically permissible when exchange term ($\bm{J}$) exceeds the on-site Coulomb repulsion ($\bm{U}$)~\cite{micnas_superconductivity_1990}. We constructed $\vec{\bm{U}}$ for Hubbard optimizattion on Sb-$5p$ ($1D$: $\vec{\bm{U}}$=$\left[U^{Sb-5p}\right]$) and both In- and Sb-$5p$ ($2D$: $\vec{\bm{U}}$=$\left[U^{In-5p},U^{Sb-5p}\right]$). This selection targets the orbitals that contribute most significantly to the states near the band edges, particularly the valence band maximum dominated by Sb-$5p$ states. While In-$4d$ semicore states are explicitly included as valence in our PAW and ONCV pseudopotentials to preserve transferability and hybridization accuracy, they lie deep in energy and do not significantly couple to the conduction or valence bands. Hence, applying a Hubbard correction to In-$4d$ states was found unnecessary and yielded negligible improvements in key observables such as band gaps or bandwidths. See Sec.~S4 of the SM for more details.
\Cref{tab:ueff} reports optimal $\bm{U_{eff}}$ values for different XC functionals with PAW and ONCV pseudopotentials.

\begin{table}[!tbp]
  \caption{Optimized effective Hubbard parameter $\bm{U_{eff}}$ of In and Sb for different pseudopotentials and XC functionals. The Hubbard correction is applied to the $5p$ orbitals of both atoms. First row's values are obtained via linear-response (LR) method. Rest of the capameters are calculated with self-authored Bayesian optimization (\cref{subsec:workflow}).}
  \label{tab:ueff}
  \begin{ruledtabular}
    \begin{tabularx}{\columnwidth}{@{} l c c c c c @{}}
      \multirow{2}{*}{\parbox{26mm}{\raggedright \textbf{Method (SOC) PP, XC func.}}} &
      \multirow{2}{*}{\parbox{14mm}{\raggedright \textbf{Optim. method}}}             &
      \multicolumn{2}{c}{\textbf{U$_\text{eff}$=$U$-$J$ (eV)}}                        &
      \multirow{2}{*}{\parbox{09mm}{\centering \textbf{E$_\text{0}$ (eV)}}}           &
      \multirow{2}{*}{\parbox{09mm}{\centering \textbf{$\bm{\Delta_\text{SO}}$ (eV)}}}                                   \\
      \cmidrule{3-4}                                                                  &    &
      \parbox{09mm}{\centering \textbf{U$_\text{eff}^\text{In-5p}$}}                  &
      \parbox{09mm}{\centering \textbf{U$_\text{eff}^\text{Sb-5p}$}}                  &    &                             \\
      \midrule
      ONCV, PBEsol                                                                    & LR & 0.85  & 2.98 & 0.070 & 0.82 \\
      PAW, PBE                                                                        & 2D & -0.20 & 3.62 & 0.240 & 0.82 \\
      ONCV, PBE                                                                       & 2D & 3.80  & 2.90 & 0.240 & 0.82 \\
      ONCV, PBEsol                                                                    & 1D & ---   & 3.91 & 0.238 & 0.83 \\
      ONCV, PBEsol                                                                    & 2D & 2.00  & 3.62 & 0.237 & 0.83 \\
    \end{tabularx}
  \end{ruledtabular}
\end{table}

\section{Results and Discussion}\label{sec:results}

\subsection{Mechanical properties: Lattice constants and Bulk Modulus}\label{subsec:mechanical_properties}

We first determine the equilibrium lattice constants and atomic positions of the zinc-blende (ZB) face-centered cubic (FCC) phase with space group \textit{F$\overline{4}$3m} for InSb. The calculated equilibrium lattice parameters and bulk moduli of InSb are summarized in \cref{tab:mechanical_electronic}. To determine the bulk modulus, we employed the Murnaghan equation of state (EOS)~\cite{murnaghan_compressibility_1944}, given by:

\begin{equation}\label{eq:eos_mur}
  E\left(V\right)=E\left(V_0\right)+\frac{B_0V}{B_0^\prime}\left[\frac{\left(\sfrac{V_0}{V}\right)^{B_0^\prime}}{B_0^\prime-1}+1\right]-\frac{B_0V_0}{B_0^\prime-1},
\end{equation}

where $E(V)$ represents the total energy as a function of volume $V$, while $E(V_0)$ corresponds to the total energy at the equilibrium volume. The bulk modulus $B_0$ quantifies the material's resistance to compression at equilibrium, and its pressure derivative $B_0^\prime$ describes the variation of this resistance with pressure. For InSb's FCC structure, the primitive unit cell volume $V$ is related to the lattice parameter $A$ by $V=(\sfrac{1}{4})A^3$. The bulk modulus and its pressure derivative are extracted by fitting the $E(V)$ data to \cref{eq:eos_mur}.

\begin{table*}[!htbp]
  \centering
  \begin{threeparttable}
    \caption{The theoretical equilibrium lattice constants $a_{eq}$ and bulk moduli $B_0$ are calculated using different pseudopotentials (with SOC) and XC functionals. The energy of the first conduction band (band gap) $E_0$ ($\Gamma_6^c - \Gamma_8^v$), the second conduction band $E_0'$ ($\Gamma_7^c - \Gamma_8^v$), the valence band spin-orbit splitting $\Delta_{SO}$ ($\Gamma_8^v - \Gamma_7^v$), and the second conduction band spin-orbit splitting $\Delta_{SO}'$ ($\Gamma_8^c - \Gamma_7^c$) are evaluated at the $\Gamma$ point. All these results are compared to previously reported values from Ref.\cite{kim_towards_2010, gant_optimally_2022} and experimental data from Ref.\cite{madelung_group_2002}. Calculations are done with $4d^{10}$ electronic states in the valence configuration of pseudopotentials unless specified.}
    \label{tab:mechanical_electronic}
    \begin{ruledtabular}
      \begin{tabularx}{\textwidth}{l c c c c c c @{}}
        \multirow{2}{*}{\parbox{55mm}{\raggedright \textbf{Methods (PP type, XC func.)}}} &
        \multicolumn{2}{c}{\textbf{Structural Properties}}                                &
        \multicolumn{4}{c}{\textbf{Electronic Properties (eV)}}                                                                                         \\
        \cmidrule{2-3} \cmidrule{4-7}                                                     &
        \parbox{12mm}{\centering \textbf{$a_{\text{eq}}$ (\AA)}}                          &
        \parbox{16mm}{\centering \textbf{$B_0$ (GPa)}}                                    &
        \parbox{10mm}{\centering \textbf{$E_0$}}                                          &
        \parbox{10mm}{\centering \textbf{$\Delta_{SO}$}}                                  &
        \parbox{10mm}{\centering \textbf{$E_0^\prime$}}                                   &
        \parbox{10mm}{\centering \textbf{$\Delta_{SO}^\prime$}}                                                                                         \\
        \midrule
        \textbf{This work: (all with SOC)}~\footnotemark[9]                               &                        &       &       &      &      &      \\
        ONCV, HSE06~\footnotemark[1]$^,$\footnotemark[2]                                  & 6.478                  & 47.95 & 0.176 & 0.84 & 3.10 & 0.46 \\
        ONCV, \HSEscreen~\footnotemark[1]$^,$\footnotemark[2]                             & 6.468                  & 48.34 & 0.232 & 0.84 & 3.15 & 0.46 \\
        ONCV, \HSEsolHFmix~\footnotemark[1]$^,$\footnotemark[2]                           & 6.465                  & 48.91 & 0.240 & 0.87 & 3.19 & 0.48 \\
        ONCV, \GoWohse~\footnotemark[1]                                                   & ---                    & ---   & 0.232 & 0.83 & 3.22 & 0.46 \\
        ONCV, PBEsol+$U$-$J$ (LR)~\footnotemark[3]                                        &                        &       & 0.070 & 0.82 & 2.87 & 0.45 \\
        PAW, PBE+$U$-$J$ ($2D$)~\footnotemark[3]                                          & 6.439                  & 53.74 & 0.241 & 0.82 & 2.99 & 0.44 \\
        ONCV, PBE+$U$-$J$ ($2D$)~\footnotemark[3]                                         & 6.431                  & 51.23 & 0.239 & 0.82 & 2.84 & 0.46 \\
        ONCV, PBESol+$U$-$J$ ($1D$)~\footnotemark[4]                                      & 6.460                  & 49.25 & 0.238 & 0.83 & 3.05 & 0.45 \\
        ONCV, PBESol+$U$-$J$ ($2D$)~\footnotemark[3]                                      & 6.468                  & 48.90 & 0.237 & 0.83 & 2.96 & 0.46 \\
        \midrule
        \textbf{Literature:}~\footnotemark[9]                                             &                        &       &       &      &      &      \\
        ONCV, \GoWo@WOT-SRSH -- Ref.\cite{gant_optimally_2022}~                           & ---                    & ---   & 0.440 & ---  & ---  & ---  \\
        PAW, HSE$_{\text{bgfit}}$ -- Ref.\cite{kim_towards_2010}~\footnotemark[5]         & 6.564                  & 42.50 & 0.240 & ---  & ---  & ---  \\
        PAW, MBJLDA$_{\text{bgfit}}$ -- Ref.\cite{kim_towards_2010}~\footnotemark[5]      & ---                    & ---   & 0.250 & ---  & ---  & ---  \\
        \midrule
        \textbf{Experiments:}                                                             &                        &       &       &      &      &      \\
        At 300~K -- Ref.\cite{madelung_group_2002}                                        & 6.479                  & 48.00 & 0.170 & 0.85 & 3.14 & 0.39 \\
        At 0~K -- Ref.\cite{madelung_group_2002}                                          & 6.469~\footnotemark[6] & ---   & 0.235 & 0.81 & 3.00 & 0.40 \\
      \end{tabularx}
    \end{ruledtabular}
    \begin{minipage}[t]{0.49\textwidth}
      \raggedright
      \footnotesize
      \footnotetext[9]{All electronic properties calculations are done with experimental lattice parameter of 6.479~\AA.}
      \footnotetext[1]{Band interpolation done with $sp^3$ projection using \Wannier.}
      \footnotetext[2]{$\bm{\alpha}$, $\bm{\mu}$ values are in \cref{tab:hse_variants}.}
      \footnotetext[3]{$^{\!,~d}$ $\bm{U_{eff}}$ values are in \cref{tab:ueff}.}
    \end{minipage}
    \hfill
    \begin{minipage}[t]{0.49\textwidth}
      \raggedright
      \footnotesize
      \footnotetext[3]{Hubbard $\bm{U}$ corrections applied to $5p$ orbitals of In and Sb.}
      \footnotetext[4]{Hubbard $\bm{U}$ corrections applied to Sb-$5p$ orbitals.}
      \footnotetext[5]{No $4d^{10}$ electrons in the valence states of pseudopotentials.}
      \footnotetext[6]{Estimated from the linear thermal expansion coefficient $\alpha = 5.37 \times 10^{-6}$ K$^{-1}$~\cite{madelung_group_2002}.}
    \end{minipage}
  \end{threeparttable}
\end{table*}

A meaningful comparison between theoretical and experimental lattice constants must account for thermal expansion effects, as experimental measurements are typically performed at finite temperatures. The experimentally reported lattice parameter of InSb at 300~K is 6.479~\AA, while its value at 0~K, estimated using the linear thermal expansion coefficient $\alpha = 5.37 \times 10^{-6}$ K$^{-1}$~\cite{madelung_group_2002}, is approximately 6.469~\AA. Thus, the theoretically predicted equilibrium lattice constants are compared with this corrected reference of 6.469~\AA.

The PAW PBE+$U$-$J$ ($2D$) and ONCV PBE+$U$-$J$ ($2D$) methods yield lattice constants of 6.439~\AA\ and 6.431~\AA, respectively, slightly underestimating the experimental 0~K reference of 6.469~\AA. The ONCV PBEsol+$U$-$J$ ($1D$ and $2D$) methods provide values of 6.460~\AA\ and 6.468~\AA, closely matching the corrected experimental value, highlighting the accuracy of the PBEsol XC functional. In comparison, the ONCV HSE06 method yields 6.478~\AA, overestimating the 0~K reference, while the ONCV \HSEscreen\ and \HSEsolHFmix\ methods give slightly underestimated values of 6.468~\AA\ and 6.465~\AA, respectively. These results indicate that PBEsol+$U$-$J$ provides an excellent balance between accuracy and efficiency, as the +$U$ corrections allow for a reliable prediction of equilibrium lattice parameters at a fraction of the computational cost of hybrid methods.

The bulk moduli $B_0$ obtained from different methods exhibit some variation: PAW PBE+$U$-$J$ ($2D$) and ONCV PBE+$U$-$J$ ($2D$) yield 53.74~GPa and 51.23~GPa, whereas ONCV PBEsol+$U$-$J$ ($2D$) and ($1D$) give 48.90~GPa and 49.25~GPa, respectively. PBE-based methods generally overestimate the bulk modulus, whereas PBEsol produces values closest to the experimental reference of 48~GPa, reinforcing its reliability for mechanical property predictions.

Since the \GW method refines quasiparticle energies rather than providing a total energy functional, it cannot be used to determine equilibrium lattice parameters and bulk moduli, which require explicit total energy minimization. Therefore, all structural predictions in this work are based on mean-field DFT calculations only.

In summary, PBEsol+$U$-$J$ predicted lattice constants and bulk modulus, guided by Bayesian optimization, achieves the best agreement with experimental ones, offering an excellent balance between accuracy and computational efficiency. This improved performance originates from the design of the PBEsol functional, which restores the second-order gradient expansion for exchange—enhancing accuracy for equilibrium properties in solids compared to the original PBE. While standard PBE+$U$ methods tend to underestimate the lattice constant and overestimate the bulk modulus, PBEsol+$U$ corrects this bias effectively. Although hybrid functionals such as HSE06 and its variants provide reasonably good predictions, they do not significantly outperform PBEsol+$U$ and come at a much higher computational cost. Therefore, PBEsol+$U$ emerges as a practical and reliable method for structural property prediction in InSb. Furthermore, the following section discusses how DFT+$U$ outperforms hybrids in predicting other fundamental energy properties and reported in figures of merit (FOMs).

\subsection{Electronic properties}\label{subsec:electronic_properties}

\subsubsection{Band structure and band gap}

The fundamental electronic properties of InSb relative to the valence band maxima (VBM), including band gaps $E_0\ ({\Gamma}_6^c-{\Gamma}_8^v)$, second conduction band energy $E_0^\prime\left(\mathrm{\Gamma}_7^c-\mathrm{\Gamma}_8^v\right)$, valence band split-off energy $\mathrm{\Delta}_{SO}\ \left(\mathrm{\Gamma}_8^v-\mathrm{\Gamma}_7^v\right)$, and second conduction band split-off energy ${\mathrm{\Delta}^\prime}_{SO}\ \left(\mathrm{\Gamma}_8^c-\mathrm{\Gamma}_7^c\right)$ are summarized in \cref{tab:mechanical_electronic}.
For all hybrid-DFT, \GoWo, and DFT+$U$ calculations, we used experimental lattice parameters and atomic coordinates due to InSb's sensitivity to geometric details.

Bulk InSb features highly localized $4d$-electrons.
Our studies show that when semicore $4d$-orbitals electrons are put into the frozen core of the constituent's pseudopotentials, the local density approximation (LDA) and GGA (such as PBE, PBEsol) XC functionals describe the band order correctly, predict the same energy of heavy-hole (HH) and light-hole (LH) states at the $\Gamma$-point but fail to determine the band gap and spin-orbit splitting accurately.
Conversely, when $4d$ states are considered as valence states, LDA, GGA, and meta-GGA-based DFT calculations yield erroneous band ordering, energy splitting between heavy hole (HH) and light hole (LH) bands, and, in extreme cases, an unphysical inverted LH band.  See Sec.~S3 of SM for more details.

In search of the answer to these unphysical effects, we analyzed the orbital-projected band structures. Near the $\Gamma$-point around the Fermi level, the $5s$ and $5p$ orbitals shape the band formation at the top of the valence band (VB) and bottom of the conduction band (CB). Even though the $4d$ orbitals exhibit negligible direct band projection near the Fermi energy region, their presence in the deep valence states induces a notable $5p$-$4d$ repulsion, significantly affecting the behavior of Sb's $5p$ electronic states. This interaction pushes the $5p$ states upwards, resulting in incorrect band ordering, HH-LH splitting, and the unphysical inversion of the LH band, ultimately leading to zero or negative band gaps. We efficiently addressed and resolved these inaccuracies with hybrid-HSE, \GW, and DFT+$U$ calculations.

\begin{figure*}[!htbp]
  \centering
  \begin{subfigure}[t]{0.325\linewidth}
    \centering
    \includegraphics[width=\textwidth]{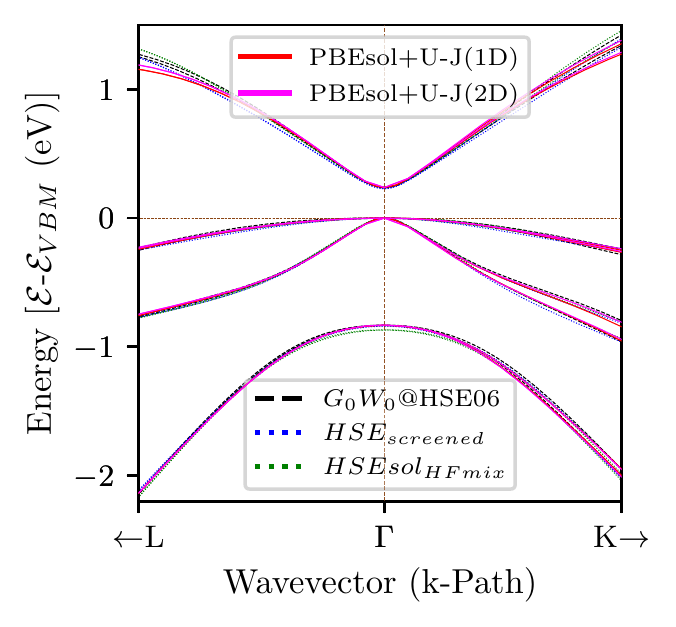}
    \caption{Compares \HSEsolHFmix, \HSEscreen, PBEsol+$U$-$J$ bands with \GoWohse.}
    \label{fig:bs_comparison}
  \end{subfigure}
  \hfill
  \begin{subfigure}[t]{0.315\linewidth}
    \centering
    \includegraphics[width=\textwidth]{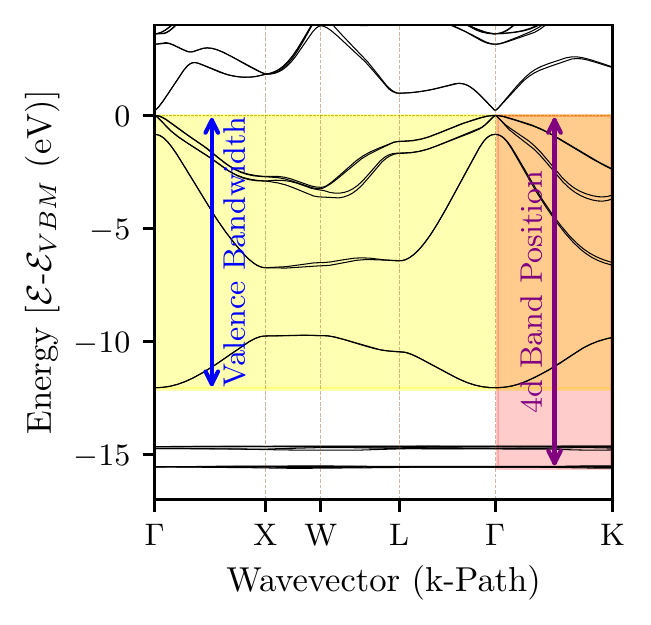}
    \caption{Valence bandwidth and highest $4d$ band position in band of \HSEscreen ($\bm{\alpha}$=0.25, $\bm{\mu}$=0.095~bohr$^{-1}$ in \cref{eq:hse_energy}).}
    \label{fig:bs_hsescreen_bw_4d}
  \end{subfigure}
  \hfill
  \begin{subfigure}[t]{0.325\linewidth}
    \centering
    \includegraphics[width=\textwidth]{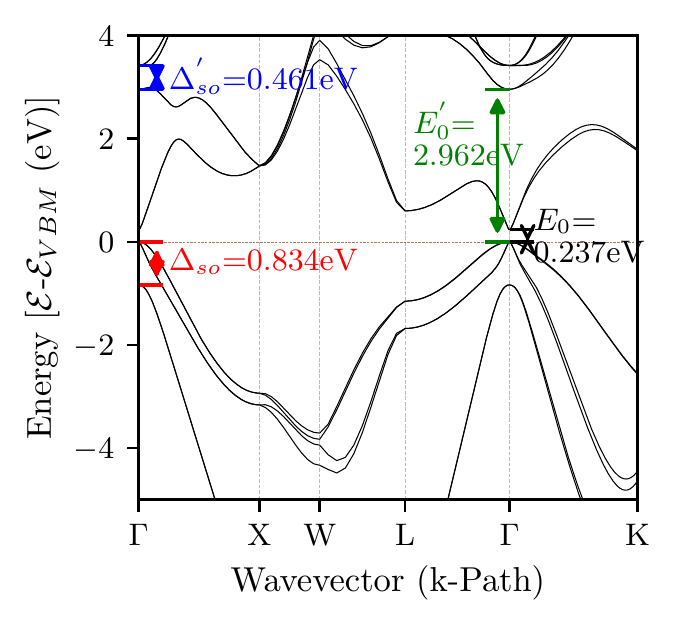}
    \caption{PBEsol+$U$-$J$ band structure: Eigen energies in \cref{eq:hub_energy} computed with $\bm{U_{eff}^{In,5p}}$=2.00~eV and $\bm{U_{eff}^{Sb,5p}}$=3.62~eV.}
    \label{fig:bs_dftu2D}
  \end{subfigure}
  \caption{InSb electronic band structures calculated with fully relativistic pseudopotentials ($4d, 5s, 5p$ valence configurations), including spin-orbit coupling, at the experimental lattice parameter (6.479~\AA).}
  \label{fig:bs_InSb_all}
\end{figure*}

The HSE06 functional, while producing a band gap of 0.18~$eV$ (underestimating the 0~$K$ experimental value of 0.23~$eV$), was refined using \GoWo on top of HSE06 (\GoWohse), yielding an accurate band gap of 0.23~$eV$, as shown in \cref{fig:bs_comparison}. Our \GoWohse with SOC marks the best improvement over previous \GoWo works, such as the study by Kim et al.~\cite{kim_towards_2010} and Gant et al.~\cite{gant_optimally_2022}, which reported overestimated band gap in the range of 0.32-0.79~$eV$. Kim et al. attributed this overestimation to the lack of self-consistent inclusion of SOC in their \GW calculations, which is crucial for materials with significant SOC. In this work, the inclusion of SOC in the calculation of both dielectric screening ($\bm{\varepsilon}$) and the self-energy ($\bm{\Sigma}$) and the incorporation of the GPP model for the inverse full-frequency dielectric matrix calculation impacted the band structure by lowering the split-off band and raising the HH and LH bands, thereby preserving the center of mass. This adjustment ensured an accurate representation of the screening properties and the band gap.

However, accurately describing the HSE band structure is essential due to the computational complexity of \GW in calculating $\bm{\varepsilon^{-1}}$ and $\bm{\Sigma}$. Even a single \GoWo iteration demands significantly more computational resources---up to two or three orders of magnitude higher than HSE calculations---depending on the $\bm{k}$ and $\bm{q}$ point grids and the number of empty states included in the evaluation of $\bm{\varepsilon}$ and $\bm{\Sigma}$. This restricts its application to systems with a few dozen atoms, highlighting the need for efficient methods like HSE to balance accuracy and resource usage. In this work, we achieved this by optimizing the range-separated screening parameter ($\bm{\mu}$) and exchange fraction ($\bm{\alpha}$). Following the optimization scheme of \HSEscreen, we obtained $\bm{\mu_{eff}}$=0.095~bohr$^{-1}$, which yields a band gap of 0.23~eV, addressing the shortcomings of HSE06 (band structure in \cref{fig:bs_hsescreen_bw_4d}). This suggests that for an accurate description of the InSb band structure, a longer short-range interaction length of about $2/\bm{\mu}_{eff}$~$\approx$~21.05~\AA\ is essential, in contrast to 18.86~\AA\ provided by HSE06. Beyond these lengths, the influence of long-range interactions predominates.

Similarly, optimizing exchange fraction $\bm{\alpha}$ in \HSEsolHFmix resulted in a band gap of 0.24~$eV$. In the HF method, the exact exchange energy cancels spurious self-interaction but does not account for electronic correlation. The exchange itself is long-ranged, decaying only as $\sfrac{1}{r}$, and not screened, leading to unrealistically high excitation energies and a considerable overestimation of the band gap. Conversely, the HSEsol functional is known for accurately treating solids but tends to underestimate the band gap. By combining HF and HSEsol in our \HSEsolHFmix functional, we achieve a balanced exchange-correlation description, resulting in a realistic and accurate band gap that fits perfectly with the 0~$K$ experimental band gap.

The band structures obtained from \HSEscreen and \HSEsolHFmix are compared with \GoWohse in \cref{fig:bs_comparison}, providing a qualitative assessment of the accuracy of the optimized $\bm{\mu}$ and $\bm{\alpha}$ parameters, respectively. A quantitative comparison is presented in \cref{tab:mechanical_electronic}. While \HSEscreen and \HSEsolHFmix align well with \GoWohse and experimental data, deviations in effective masses and Luttinger parameters (see \cref{tab:m_eff}, \cref{tab:luttinger}) arise from the highly localized $4d$-orbital electron density ($\sim$ 15 eV in \cref{fig:bs_hsescreen_bw_4d}) and self-interaction errors (SIE) inherent in standard hybrid functionals.

These limitations are addressed by Hubbard-corrected DFT+$U$-$J$ (see \cref{subsubsec:dftu}), which reduces SIE via piecewise linearity in the energy functional. The PBEsol+$U$-$J$ band structure, obtained from a two-dimensional ($2D$) $U_{eff}$ optimization, is shown in \cref{fig:bs_dftu2D}. Strong agreement of band parameters, effective masses, and Luttinger parameters from four different DFT+$U$-$J$ calculation with experimental values underscores the computational efficiency of DFT+$U$-$J$ over hybrid functionals and \GoWo in electronic structure predictions.
A detailed comparative analysis of the computational costs for \GoWo, hybrid functionals, and DFT+$U$-$J$ methods, including scaling behavior, is provided in the SM (Sec.~S6).

\begin{table*}[!tbp]
  \caption{Comparative analysis of valence bandwidths and 4d band positions for InSb using various computational methods. The valence bandwidth is defined as the maximal energy difference between the top four (excluding spin degeneracy) valence bands. The highest $4d$ band position is defined as the highest $4d$ orbital absolute eigen energies relative to the valence band maxima. The table also presents the mean absolute relative error (MARE) with respect to the leftmost experimental value presented, providing a quantitative measure of their performance.}
  \label{tab:valence_bw_4d_pos}
  \begin{ruledtabular}
    \begin{tabularx}{\textwidth}{@{} l c c c c @{}}
      \multirow{2}{*}{\parbox{45mm}{\raggedright \textbf{Methods (PP type, XC func.)}}} &
      \multicolumn{2}{c}{\textbf{Valence bandwidths}}                                   &
      \multicolumn{2}{c}{\textbf{Highest 4d band positions}}                                                                                                                                                                                                                                                                              \\
      \cmidrule{2-3} \cmidrule{4-5}                                                     &
      \parbox{27mm}{\centering\textbf{in eV}}                                           &
      \parbox{27mm}{\centering\textbf{MARE}}                                            &
      \parbox{29mm}{\centering\textbf{in eV (absolute)}}                                &
      \parbox{29mm}{\centering\textbf{MARE}}                                                                                                                                                                                                                                                                                              \\
      \midrule
      \textbf{This work (all with SOC)}                                                 &                                                                                                                            &                                                                                                   &       &        \\
      ONCV, HSE06                                                                       & 12.10                                                                                                                      & 0.1204                                                                                            & 15.70 & 0.0754 \\
      ONCV, \HSEscreen                                                                  & 12.10                                                                                                                      & 0.1204                                                                                            & 15.60 & 0.0813 \\
      ONCV, \HSEsolHFmix                                                                & 12.27                                                                                                                      & 0.1361                                                                                            & 15.71 & 0.0748 \\
      ONCV, \GoWohse                                                                    & 11.34                                                                                                                      & 0.0500                                                                                            & 16.56 & 0.0247 \\
      PAW, PBE+$U$-$J$ ($2D$)                                                           & 10.52                                                                                                                      & 0.0259                                                                                            & 15.20 & 0.1048 \\
      ONCV, PBE+$U$-$J$ ($2D$)                                                          & 10.75                                                                                                                      & 0.0046                                                                                            & 14.93 & 0.1207 \\
      ONCV, PBEsol+$U$-$J$ ($1D$)                                                       & 10.33                                                                                                                      & 0.0435                                                                                            & 14.99 & 0.1172 \\
      ONCV, PBEsol+$U$-$J$ ($2D$)                                                       & 10.83                                                                                                                      & 0.0028                                                                                            & 14.91 & 0.1219 \\
      \midrule
      \textbf{Literature:}                                                              &                                                                                                                            &                                                                                                   &       &        \\
      WOT-SRSH -- Ref.\cite{gant_optimally_2022}                                        & 11.96                                                                                                                      & 0.1074                                                                                            & 16.24 & 0.0436 \\
      \GoWo@WOT-SRSH -- Ref.\cite{gant_optimally_2022}                                  & 11.32                                                                                                                      & 0.0481                                                                                            & 16.55 & 0.0253 \\
      \midrule
      \textbf{Experiments:}                                                             & \multicolumn{2}{c}{10.8 (ARPES)~\cite{middelmann_valence-band-structure_1986}, 11.7 (XPS)~\cite{goldmann_electronic_1989}} & \multicolumn{2}{c}{16.98~\cite{cardona_photoemission_1972}, 17.1~\cite{goldmann_electronic_1989}}                  \\
    \end{tabularx}
  \end{ruledtabular}
\end{table*}

\subsubsection{Valence bandwidths}

We define the valence bandwidth as the maximal energy difference between the top four valence bands (see \cref{fig:bs_hsescreen_bw_4d}), excluding spin degeneracy, which is characteristic of ZB materials with strong $sp^3$ hybridization. The calculated valence bandwidths for InSb are summarized in \cref{tab:valence_bw_4d_pos}.

ONCV HSE06 and \HSEscreen yield a bandwidth of 12.1~eV, indicating that screening effects have minimal impact on the bandwidth. The ONCV \HSEsolHFmix slightly increases it to 12.27~eV, suggesting improved treatment of electronic interactions. The ONCV \GoWohse method results in a reduced bandwidth of 11.34~eV, reflecting many-body effects.

DFT+$U$ methods predict lower bandwidths compared to hybrid and \GoWo methods: PAW PBE+$U$-$J$ ($2D$) at 10.52~eV, ONCV PBE+$U$-$J$ ($2D$) at 10.75~eV, ONCV PBEsol+$U$-$J$ ($2D$) at 10.83~eV, and ONCV PBEsol+$U$-$J$ ($1D$) at 10.33~eV.

To assess accuracy, the mean absolute relative error (MARE) is computed for each method relative to the experimental ARPES bandwidth of 10.8~eV~\cite{middelmann_valence-band-structure_1986}. Among DFT+$U$ methods, ONCV PBEsol+$U$-$J$ ($2D$) achieves the lowest MARE, closely matching experiment. Both \GoWohse and DFT+$U$ methods demonstrate similar accuracy in predicting valence bandwidths. Additionally, our DFT+$U$ results surpass literature values from WOT-SRSH (11.96~eV) and \GoWo@WOT-SRSH (11.32~eV)~\cite{gant_optimally_2022}, further validating their reliability.

\subsubsection{4$d$ band energies}

We define the $4d$ band position as the absolute eigenvalue of the deepest $4d$ orbital relative to the VBM, as shown in \cref{fig:bs_hsescreen_bw_4d}. The highest $4d$ band positions for InSb are summarized in \cref{tab:valence_bw_4d_pos}. Experimentally, this value is 16.98~eV~\cite{cardona_photoemission_1972}. Although the $4d$ states correspond to negative binding energies, they are reported as absolute values in the table for clarity. The mean absolute relative error (MARE) is used to assess the accuracy of each method.

In InSb, the Sb $4d$ states primarily determine the $4d$ band position, with In $4d$ states lying approximately 1~eV higher. DFT+$U$ methods predict Sb-$4d$ states around 15~eV and In-$4d$ states around 14~eV. The ONCV HSE06 and ONCV \HSEscreen methods yield 15.7~eV and 15.6~eV, respectively, slightly underestimating experiment. The ONCV \HSEsolHFmix method gives 15.71~eV, aligning with ONCV HSE06.

The ONCV \GoWohse method provides the closest agreement to experiment at 16.56~eV, with the lowest MARE of 0.0247, demonstrating the accuracy of \GW corrections. In contrast, DFT+$U$ methods systematically underestimate the $4d$ positions, with PAW PBE+$U$-$J$ ($2D$) at 15.2~eV, and ONCV PBE+$U$-$J$ ($2D$) and ONCV PBEsol+$U$-$J$ ($2D$) yielding 14.93~eV and 14.91~eV, respectively. The ONCV PBEsol+$U$-$J$ ($1D$) method gives 14.99~eV. While DFT+$U$ offers computational efficiency, its prediction accuracy for $4d$ positions remain slightly lower than experimental values.

Notably, the ONCV \GoWohse method (16.56~eV) aligns well with literature values, such as 16.24~eV from WOT-SRSH and 16.55~eV from \GoWo@WOT-SRSH~\cite{gant_optimally_2022}, reinforcing the effectiveness of \GW corrections for accurate $4d$ band energy predictions. Overall, \GW-corrected methods show the best agreement with experiment, followed by hybrid functionals, whereas DFT+$U$ methods, despite their efficiency, yield slightly lower $4d$ positions.

\subsection{Figure of merits}\label{subsec:fom}

\subsubsection{Effective electron and hole Masses}

\begin{table*}[!tbp]
  \caption{Effective masses ($|m^\ast/m_e|$) of heavy-hole (HH), light-hole (LH), split-off (SO), and first conduction electron (Elec.) bands in InSb at $\Gamma$-point along various crystallographic directions ($\Gamma$-X [100], $\Gamma$-K [110], $\Gamma$-L [111]). The data includes results from different methods, literature values, and experimental measurements. Mean absolute error (MAE) and root mean squared error (RMSE) are provided to quantify the performance of each method.}
  \label{tab:m_eff}
  \begin{ruledtabular}
    \begin{tabularx}{\textwidth}{@{} l c c c c c c c @{}}
      \textbf{Methods}                                                 &
      \textbf{Direction}                                               &
      \textbf{$|\sfrac{\text{m}_\text{SO}^*}{\text{m}_\text{e}}|$}     &
      \textbf{$|\sfrac{\text{m}_\text{LH}^*}{\text{m}_\text{e}}|$}     &
      \textbf{$|\sfrac{\text{m}_\text{HH}^*}{\text{m}_\text{e}}|$}     &
      \textbf{$|\sfrac{\text{m}_\text{Elec.}^*}{\text{m}_\text{e}}|$}  &
      \textbf{MAE}                                                     &
      \textbf{RMSE}                                                                                                                                                            \\
      \midrule
      ONCV HSE06                                                       & $\Gamma$~--~X~[100] & 0.125 & 0.020 & 0.235 & 0.021 & \multirow{3}{*}{0.016} & \multirow{3}{*}{0.021} \\
                                                                       & $\Gamma$~--~K~[110] & 0.129 & 0.021 & 0.457 & 0.020 &                        &                        \\
                                                                       & $\Gamma$~--~L~[111] & 0.129 & 0.022 & 0.608 & 0.022 &                        &                        \\
      ONCV, \HSEscreen                                                 & $\Gamma$~--~X~[100] & 0.118 & 0.020 & 0.248 & 0.022 & \multirow{3}{*}{0.015} & \multirow{3}{*}{0.022} \\
                                                                       & $\Gamma$~--~K~[110] & 0.129 & 0.019 & 0.457 & 0.019 &                        &                        \\
                                                                       & $\Gamma$~--~L~[111] & 0.129 & 0.021 & 0.618 & 0.023 &                        &                        \\
      ONCV, \HSEsolHFmix                                               & $\Gamma$~--~X~[100] & 0.121 & 0.021 & 0.245 & 0.021 & \multirow{3}{*}{0.018} & \multirow{3}{*}{0.029} \\
                                                                       & $\Gamma$~--~K~[110] & 0.123 & 0.021 & 0.462 & 0.018 &                        &                        \\
                                                                       & $\Gamma$~--~L~[111] & 0.126 & 0.023 & 0.645 & 0.025 &                        &                        \\
      ONCV, \GoWohse                                                   & $\Gamma$~--~X~[100] & 0.105 & 0.021 & 0.266 & 0.019 & \multirow{3}{*}{0.010} & \multirow{3}{*}{0.018} \\
                                                                       & $\Gamma$~--~K~[110] & 0.112 & 0.021 & 0.445 & 0.021 &                        &                        \\
                                                                       & $\Gamma$~--~L~[111] & 0.112 & 0.019 & 0.617 & 0.019 &                        &                        \\
      PAW, PBE+$U$-$J$ ($2D$)                                          & $\Gamma$~--~X~[100] & 0.127 & 0.017 & 0.272 & 0.019 & \multirow{3}{*}{0.024} & \multirow{3}{*}{0.048} \\
                                                                       & $\Gamma$~--~K~[110] & 0.147 & 0.013 & 0.454 & 0.017 &                        &                        \\
                                                                       & $\Gamma$~--~L~[111] & 0.142 & 0.016 & 0.712 & 0.018 &                        &                        \\
      ONCV, PBE+$U$-$J$ ($2D$)                                         & $\Gamma$~--~X~[100] & 0.125 & 0.018 & 0.244 & 0.018 & \multirow{3}{*}{0.017} & \multirow{3}{*}{0.030} \\
                                                                       & $\Gamma$~--~K~[110] & 0.140 & 0.015 & 0.427 & 0.014 &                        &                        \\
                                                                       & $\Gamma$~--~L~[111] & 0.140 & 0.017 & 0.646 & 0.016 &                        &                        \\
      ONCV, PBEsol+$U$-$J$ ($1D$)                                      & $\Gamma$~--~X~[100] & 0.114 & 0.019 & 0.249 & 0.015 & \multirow{3}{*}{0.012} & \multirow{3}{*}{0.029} \\
                                                                       & $\Gamma$~--~K~[110] & 0.106 & 0.015 & 0.448 & 0.015 &                        &                        \\
                                                                       & $\Gamma$~--~L~[111] & 0.106 & 0.014 & 0.656 & 0.014 &                        &                        \\
      ONCV, PBEsol+$U$-$J$ ($2D$)                                      & $\Gamma$~--~X~[100] & 0.119 & 0.016 & 0.255 & 0.016 & \multirow{3}{*}{0.013} & \multirow{3}{*}{0.035} \\
                                                                       & $\Gamma$~--~K~[110] & 0.111 & 0.018 & 0.426 & 0.014 &                        &                        \\
                                                                       & $\Gamma$~--~L~[111] & 0.111 & 0.017 & 0.675 & 0.015 &                        &                        \\
      \midrule
      \textbf{Literature:}                                             &                     &       &       &       &       &                        &                        \\
      PAW, HSE$_{\text{bgfit}}$ -- Ref.\cite{kim_towards_2010}         & $\Gamma$~--~X~[100] & 0.129 & 0.018 & 0.245 & 0.022 & 0.030                  & 0.053                  \\
      PAW, MBJLDA$_{\text{bgfit}}$ -- Ref.\cite{kim_towards_2010}      & $\Gamma$~--~X~[100] & 0.150 & 0.024 & 0.292 & 0.017 & 0.029                  & 0.037                  \\
      \midrule
      \textbf{Experiments:} at 300~K -- Ref.\cite{madelung_group_2002} & $\Gamma$~--~X~[100] & 0.111 & 0.015 & 0.263 & 0.014 &                        &                        \\
                                                                       & $\Gamma$~--~K~[110] & 0.111 & 0.015 & 0.435 & 0.014 &                        &                        \\
                                                                       & $\Gamma$~--~L~[111] & 0.110 & 0.014 & 0.556 & 0.014 &                        &                        \\
    \end{tabularx}
  \end{ruledtabular}
\end{table*}

In zinc-blende InSb, the inherent symmetries result in consistent split-off and electron masses across the [100] ($\Gamma$-$X$), [110] ($\Gamma$-$K$), and [111] ($\Gamma$-$L$) directions.
However, the effective masses of light-hole and heavy-hole vary significantly along these directions.
This variation arises from the asymmetric and nonparabolic nature of the electronic bands around the $\Gamma$-point, necessitating a nonparabolic fitting approach.
This method is essential because the electronic bands deviate from a simple quadratic dispersion near the $\Gamma$-point.
The relationship between the energy eigenvalues $\mathcal{E}$, the wave vector $\bm{k}$, and the effective mass $m_{eff}$ is given by \cref{eq:m_eff}.

\begin{equation}\label{eq:m_eff}
  \begin{matrix}
    \mathcal{E}+\alpha\cdot\mathcal{E}^2=c_0+c_1k+c_2k^2 \\
    m_{eff}=\frac{\hbar^2k^2}{2c_2}                      \\
  \end{matrix}
\end{equation}

Here, $\hbar$ is the reduced Planck constant, and $\hbar^2 k^2$ represents the kinetic energy associated with the particle's wave-like behavior. This fitting method accounts for higher-order terms $\alpha \mathcal{E}^2$ that reflect the nonparabolic nature of the bands.
The variations in the effective masses across different crystallographic directions ([100], [110], and [111]) are indicative of the degree of nonparabolicity ($\alpha$), particularly in the heavy-hole (HHh) band.
This approach ensures an accurate determination of effective masses, capturing the true nature of the electronic bands in InSb, where simple parabolic fits would be insufficient.

The results for the effective electron and hole carrier masses are summarized in \cref{tab:m_eff} and compared to experimental values from Ref.\cite{madelung_group_2002}, along with the calculated mean absolute error (MAE) and root-mean-squared error (RMSE).
The ONCV HSE06 method predicts electron and hole masses relatively close to experimental values, with an MAE of 0.016 and an RMSE of 0.021.
The ONCV \HSEsolHFmix method slightly overestimates the split-off, light-hole, and electron band effective masses in all directions, resulting in an MAE of 0.018 and an RMSE of 0.029.
The ONCV \HSEscreen method shows modest improvements, with effective masses that are generally closer to experimental values than those obtained with the ONCV HSE06 and \HSEsolHFmix method, reflected in an MAE of 0.015 and an RMSE of 0.022.
We have also seen that all methods underestimate the heavy-hole mass in [100] direction.
For example, the heavy-hole mass along the [100] direction is 0.248, compared to the experimental value of 0.263.

The ONCV \GoWohse method provides the most accurate results among the tested methods, with an MAE of 0.010 and an RMSE of 0.018.
This highlights the importance of including many-body effects in electronic structure calculations.
In contrast, DFT+$U$ methods, while computationally much more efficient than either of the HSE-based and \GW methods, show a more comprehensive range of results.
For instance, the PAW PBE+$U$-$J$ ($2D$) method tends to overestimate the heavy-hole mass, yielding an MAE of 0.024 and an RMSE of 0.048.
Similarly, the ONCV PBE+$U$-$J$ ($2D$) method shows an MAE of 0.017 and an RMSE of 0.030, slightly predicting a better effective result than PAW methods.
The ONCV PBEsol+$U$-$J$ $1D$ and $2D$ methods result in an MAE of 0.012 and 0.013 with an RMSE of 0.029 and 0.035, respectively, predicting the second-best effective mass results after \GW and significantly closer to the experimental value.
This also suggests the better performance of PBEsol XC functionals over PBE functionals in our calculations.

Comparing our calculated effective masses with literature and experimental values, we find that hybrid functionals (ONCV HSE06, ONCV \HSEscreen, and ONCV \HSEsolHFmix) and the \GW corrected method (ONCV \GoWohse) provide high accuracy at increased computational cost.
While \GW methods offer superior precision, their expense necessitates a trade-off between accuracy and efficiency in electronic structure calculations.
Among all methods, PBEsol+$U$-$J$ achieves the best balance, particularly in $1D$ and $2D$ cases, making it the most suitable choice for effective mass calculations.

\subsubsection{Luttinger parameters}

\begin{table}[!tbp]
  \caption{Luttinger parameters ($\gamma_1$, $\gamma_2$, $\gamma_3$) of InSb calculated using various methods, compared against literature and experimental values. Root mean squared absolute relative errors (RMSARE) are included to evaluate the performance of each method.}
  \label{tab:luttinger}
  \begin{ruledtabular}
    \begin{tabularx}{\textwidth}{@{} l c c c c @{}}
      \textbf{Method}                                        &
      \textbf{$\gamma_{\text{1}}$}                           &
      \textbf{$\gamma_{\text{2}}$}                           &
      \textbf{$\gamma_{\text{3}}$}                           &
      \textbf{RMSARE}                                                                        \\
      \midrule
      ONCV, HSE06                                            & 27.13 & 11.62 & 12.74 & 0.233 \\
      ONCV, \HSEscreen                                       & 27.02 & 11.52 & 12.70 & 0.237 \\
      ONCV, \HSEsolHFmix                                     & 25.85 & 10.87 & 12.15 & 0.274 \\
      ONCV, \GoWohse                                         & 29.69 & 11.91 & 12.03 & 0.223 \\
      PAW, PBE+$U$ ($2D$)                                    & 31.25 & 13.25 & 14.92 & 0.116 \\
      ONCV, PBE+$U$ ($2D$)                                   & 29.83 & 12.48 & 14.14 & 0.162 \\
      ONCV, PBEsol+$U$ ($1D$)                                & 28.32 & 11.92 & 13.40 & 0.203 \\
      ONCV, PBEsol+$U$ ($2D$)                                & 33.21 & 14.05 & 15.86 & 0.064 \\
      \midrule
      \textbf{Literature:}                                   &       &       &       &       \\
      HSE$_{\text{bgfit}}$ -- Ref.\cite{kim_towards_2010}    & 29.44 & 12.79 & 13.85 & 0.163 \\
      MBJLDA$_{\text{bgfit}}$ -- Ref.\cite{kim_towards_2010} & 22.26 & 9.47  & 10.31 & 0.375 \\
      \midrule
      \textbf{Experiments:}                                  &       &       &       &       \\
      At 300~K -- Ref.\cite{madelung_group_2002}             & 34.80 & 15.50 & 16.50 & 0     \\
      At 300~K -- Ref.\cite{madelung_group_2002}             & 35.10 & 15.60 & 16.70 &       \\
    \end{tabularx}
  \end{ruledtabular}
\end{table}

The Luttinger parameters are crucial for characterizing the valence band and providing insights into the electronic properties of semiconductor materials.
The topology of the valence bands, described by these three Luttinger parameters, directly influences the effective conduction-band masses, which are critical for accurately predicting the behavior of charge carriers in semiconductors.
Thus, determining Luttinger parameters is critical to designing and analyzing electronic and optoelectronic devices.
These Luttinger parameters $\gamma_1$, $\gamma_2$, and $\gamma_3$ were derived from effective masses obtained via DFT band structure calculations, as explicitly calculated in Ref.\cite{madelung_group_2002}.
This work employed the least-square method to resolve the complex and non-linear relationships of effective masses in the Luttinger parameter equations.
The accuracy of the calculated Luttinger parameters, listed in \cref{tab:luttinger}, was validated by comparing them with reported literature values and experimental data using the root mean squared absolute relative error (RMSARE), showing excellent agreement and validating our computational approach.

Among the methods, the hybrid functional approaches, such as ONCV HSE and its variants, provide a reasonable accuracy.
For instance, ONCV \HSEsolHFmix yields a relatively high RMSARE of 0.274.
The screened variant of this method, ONCV \HSEscreen, shows better performance metrics with slightly different Luttinger parameters.
The ONCV \GoWohse method stands out among the HSE and \GW methods with an RMSARE of 0.223, indicating its effectiveness in predicting Luttinger parameters, band energies, and effective masses.
However, the computational expense of \GW methods often limits their practical use.

Interestingly, methods incorporating the Hubbard $\bm{U}$ correction exhibit an excellent balance between computational cost and accuracy, like those seen in effective masses.
Notably, the ONCV PBEsol+$U$-$J$ ($2D$) method shows the lowest RMSARE of 0.064, closely aligning with experimental values.
This method outperforms \GoWo (0.223 RMSARE) and HSE-based approaches (0.233-0.274 RMSARE), and the results reported in the literature, such as the PAW HSE$_\text{bgfit}$ (0.163 RMSARE) and MBJLDA$_\text{bgfit}$ (0.375 RMSARE) methods as reported in \cref{tab:luttinger}.

\subsection{Transferability of Optimized Parameters}\label{subsec:transferability}

To validate the robustness and transferability of our Bayesian-optimized parameters, specifically the Hubbard parameter $\bm{U_{eff}}$ and the hybrid HSEsol functional's HF exchange fraction parameter $\bm{\alpha}$, we tested their predictive performance under hydrostatic strain conditions. Hydrostatic strain maintains cubic ZB symmetry but significantly alters lattice parameters, electronic screening, and orbital hybridization, thus providing a rigorous test of the optimized parameters outside their fitting conditions.

\begin{figure}[!tbp]
  \centering
  \includegraphics[width=\columnwidth]{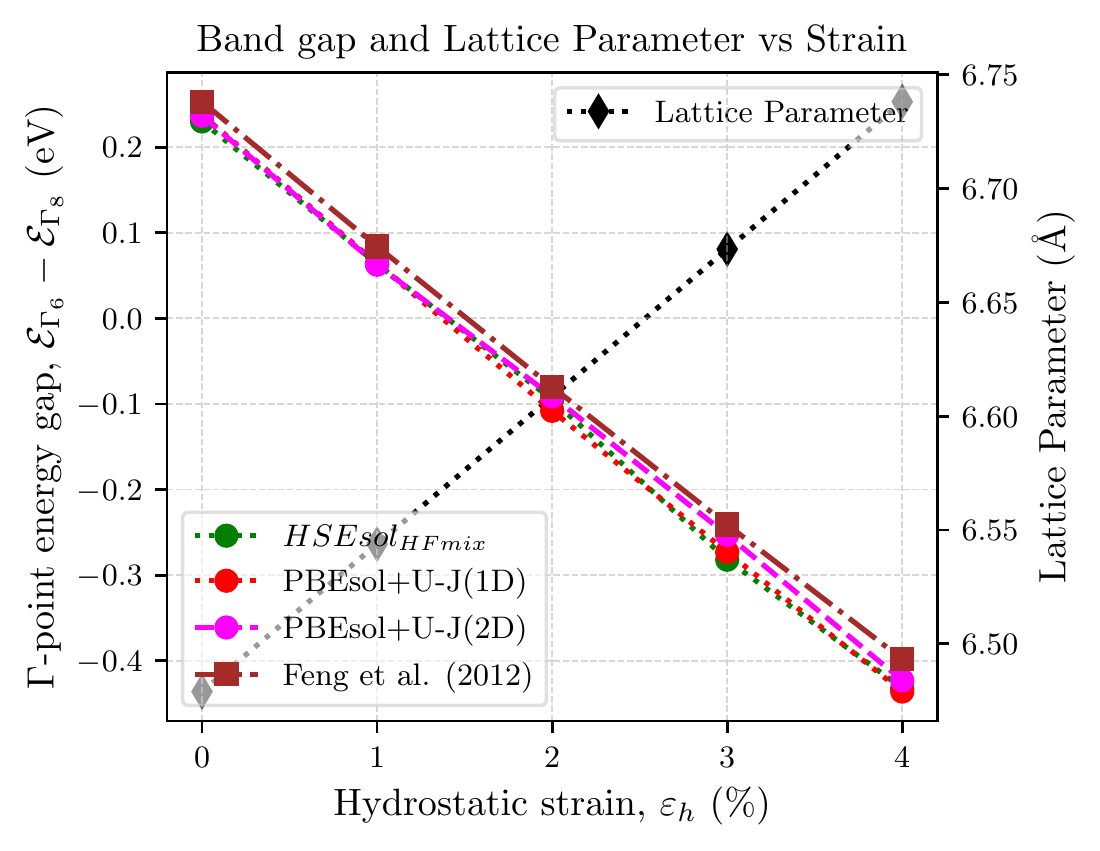}
  \caption{$\Gamma$-point band gap ($E_{\Gamma,6^c} - E_{\Gamma,8^v}$) in InSb under hydrostatic strain, calculated using \HSEsolHFmix and PBEsol+$U$-$J$ with $\bm{U_{eff}}$ and $\bm{\alpha}$ values transferred from unstrained bulk optimized parameters in \cref{tab:hse_variants} and \cref{tab:ueff}, respectively. The results, obtained without re-fitting, are consistent with prior DFT trends~\cite{feng_strain_2012}, confirming the predictive robustness of the optimized parameters.}
  \label{fig:bandgap_strain}
\end{figure}

Figure~\ref{fig:bandgap_strain} presents the evolution of the direct $\Gamma$-point band gap ($E_{\Gamma,6^c}-E_{\Gamma,8^v}$) calculated by transferring the optimized $\bm{\alpha}$ parameter reported in \cref{tab:hse_variants} (for \HSEsolHFmix) and the optimized $\bm{U_{eff}}$ values reported in \cref{tab:ueff} (for PBEsol+$U$-$J$ methods) to hydrostatically strained InSb structures. All methods with these transferred parameters show consistent behavior: compressive strain ($\varepsilon_h < 0$) increases the band gap due to enhanced orbital hybridization, while tensile strain ($\varepsilon_h > 0$) reduces it, eventually closing the gap around $\varepsilon_h \approx +1\%$ and causing a full band inversion at $\varepsilon_h = +2\%$. These results are in excellent agreement with previous theoretical reports~\cite{feng_strain_2012}, clearly demonstrating the transferability of the optimized parameters beyond their equilibrium fitting conditions.

Further evidence of parameter transferability, including detailed projected band structures verifying correct orbital contributions and capturing the strain-induced band inversion mechanism, is provided in the SM (Sec.~S5, Fig.~S2).

Additional validation of these parameters in alloyed systems (\InAsSb) is presented in a separate manuscript currently under preparation.

\section{Conclusions}\label{sec:conclusion}

This investigation into the electronic properties of InSb through advanced \textit{ab initio} methodologies has led to significant strides in semiconductor physics.
By explicitly including the often-neglected semicore $4d^{10}$ electrons--particularly those of In atoms--as valence states, using fully relativistic PAW and ONCV pseudopotentials, we overcome longstanding limitations in conventional DFT modeling in \textit{prior} literature, such as non-physical band inversion, underestimated or vanishing band gaps, and achieve a more accurate description of inSb's electronic structure.

By employing a synergy of sophisticated methods---including the hybrid-HSE functional, \GoWo, and DFT+$U$---we have significantly enhanced the qualitative and quantitative accuracy of band structure predictions.
The self-consistent inclusion of spin-orbit coupling (SOC) in \GoWo calculations, starting from HSE06 mean-field wavefunctions (\GoWohse), corrects the band gap overestimations observed in previous studies and yields gap predictions in excellent agreement with experimental values. This highlights the critical importance of SOC in materials with strong spin-orbit interactions.

Leveraging a Bayesian optimization framework, we systematically tuned three key parameters--the inverse screening length $\bm{\mu}$ in \HSEscreen, the exchange fraction $\bm{\alpha}$ in \HSEsolHFmix, and the Hubbard $\bm{U}$ in DFT+$U$--to significantly reduce computational cost relative to the reference \GoWohse method, while maintaining high fidelity in electronic structure predictions.

Our optimized approaches demonstrate excellent agreement with experimental data across multiple physical quantities, including the electronic band gap, valence bandwidth, $4d$ semicore band position, effective masses, and Luttinger parameters. The reference \GoWohse calculation yields a direct band gap error below 1.3\% relative to the experimental 0~K value of 0.235~eV, along with a valence bandwidth mean absolute relative error (MARE) of 0.050, a $4d$ band position MARE of 0.024, effective mass mean absolute error (MAE) of 0.010, and a Luttinger parameter root mean squared absolute relative error (RMSARE) of 0.22.

All Bayesian-optimized hybrid-HSE and DFT+$U$ methods accurately reproduce the experimental band gap, with deviations less than 5~meV. For valence bandwidths, DFT+$U$ methods exhibit excellent accuracy with MARE below 0.04, while HSE-based methods retain slightly higher but still good accuracy with MARE below 0.14. In predicting $4d$ band positions, HSE-based methods perform better (MAREs $<$ 0.08) compared to DFT+$U$ (MAREs $\approx$ 0.11-0.12). For effective masses, DFT+$U$ methods achieve MAEs below 0.025, while HSE-based methods attain slightly better accuracy with MAEs below 0.020. Notably, Luttinger parameters are best captured by 2D-BO-optimized DFT+$U$ (RMSAREs $<$ 0.2), surpassing HSE-based results (RMSAREs $<$ 0.3).

These results underscore the predictive accuracy, efficiency, and transferability of our BO-enhanced framework, affirming its suitability for cost-effective modeling of narrow-gap, spin-orbit coupled semiconductors such as InSb.

Our integration of \textit{ab initio} methods---\GoWohse, hybrid-HSE-based methods, and DFT(PBE, PBEsol)+$U$---with Bayesian optimization establishes a robust and generalizable approach to electronic structure prediction.

Beyond InSb, the transferability of the optimized parameters has been rigorously validated under hydrostatic strain, and will be further extended in a forthcoming study to \InAsSb alloys and structurally perturbed systems. These applications aim to demonstrate the framework's broader utility in band-structure engineering and materials design across technologically relevant III-V and II-VI semiconductors.

\begin{acknowledgments}
  The computational work for this study was performed on the ``Ruche'' supercomputing cluster at the Institute for Development and Resources in Intensive Scientific Computing (IDRIS), CNRS, Université Paris-Saclay.
  This research was supported by the grant from the French Ministry of Higher Education, Research and Innovation (MESRI, France).
\end{acknowledgments}

\end{document}